# A Survivability Strategy in route optimization Mobile Network by memetic algorithm

K .K. Gautam
Department of Computer Science & Engineering
Roorkee Engineering & Management Technology Institute,
Shamli (247774) India

Anurag Rai
Department of Information Technology
College of Engineering Roorkee
Roorkee (247667) India

*Abstract*— **The capability to provide network service even under a significant network system element disruption is the backbone for the survival of route optimize of mobile network Technology in today's world. Keeping this view in mind, the present paper highlights a new method based on memetic algorithm .**

I. INTRODUCTION

Network survivability is considered to cope, it with increasing demand for reliable network system. Network survivability for the route optimization is an essential aspect of reliable communication service. Survivability consists not only of robustness against failure occurring due to natural faults. In mobile networks infrastructure element such as base station (BS), base station Controller (BSC), wired links, and mobile switch centre (MSC), are employed to provide and maintain essential services, hence the operation interruption of a network component affects overall or partial network services. Wireless access network have unique characteristics to support mobile users, which can result in different survivability and security aspect [3]. Therefore wireless survivability strategies must be designed to improve the service available rate of the network component system (4), (3).

Mobile user authentication is necessary when a mobile user wants to request service provided by the service providers survivable (SPS) in the visited domains. In this paper, we present a survivability strategy in mobile networks method by the use of memetic algorithm.

A network could be as simple as a forum held in a city between people, where people use the opportunity to communicate with each other, they use a network by the use of memetic algorithm has the potential for setup the survivability. Fundamental to distribute mechanics is the effect of measurement on a state. If some property of a general state is measured, it collapses to an eigenstate of the property and cannot be 'rebuilt' into the original state. Information can be encoded into a general optimize set up. Mobile networks can have very complex form of hierarchy e.g. Mobile networks in a mobile network visiting mobile nodes (VMNS) in mobile networks and so on. This situation is called as a nested mobile network.

Many important problems arising in science, industry and commerce, mobile networks fall very neatly into the read-made category of optimization problem.

II. SURVIVABILITY

Traditional security research is primarily focused on the detection and prevention of intrusion and attacks rather than on continued correct operation while under attack. Fault tolerance is usually concerned with redundancy that is required to detect and correct up to a given number of naturally occurring faults. Nature is not malicious and conventional failure model make significant assumptions, in particular, assuming faults to be independent and random. The presence of intelligent adversarial attacks can protocol vulnerability often become more important considerations in the presence of an adversary[1,2]

There are a number of definitions of survivability. The one we use here is from the Software Engineering Institute, which emphasizes timeliness, survivability under attack and failure, and that detection of attack is a vital capability.

Survivability is the capability of a system to fulfill its mission in a timely manner, even in the presence of attacks or failures. Survivability goes into the demon of security and fault. Tolerance of focus on delivery of essential service even when system is entered or experiences failures, and rapid recovery of full service, when conditions improve. Unlike traditional security measures that require central control and administrative, survivability addresses highly distributed unbounded network environment that lacks central control and unified security policies. Mobile networks can have very complex form of hierarchy e.g. Mobile networks in a mobile network visiting mobile nodes (VMNS) in mobile networks and so on.

### III. THE THREE RS: RESISTANCE, RRCOGNITION, AND RECOVERY

The focus of survivability is on delivery of essential services and preservation of essential assets. Essential service and asserts are those system capabilities that are critical o fulfilling mission objectives. Survivability depends on three key capabilities: resistance, recognition, and recovery. Resistance is the capability to detect attacks as they occur and to evaluate the extent of damage and compromise. Recovery, a hallmark of survivability is the capability to maintain essential service and asserts during attacks, limit the extent of damage and restore full service following attack.

We further extend this definition to require that survivability system be able to quickly incorporate lesion learned from failure, evolve, and adapt to emerging threats. We call this survivability feature refinement.

We can classify survivable mobile wireless networking requirement into four categories based on (2): (i) resistance requirement; (ii) recognition requirement; (iii) recovery requirements; and (iv) refinement requirement. We can also describe a requirement definition process (3). This includes the definition of system and survivability requirement, legitimate and intruder usage requirement, development requirement, operation requirement, and evolution requirement. Essential service must be identified and specified for the penetration, exploration, and exploitation phases of the attack.

The approach has guided this work and is recommended for more detailed requirement analyses for future mobile wireless network.

Ultimately, there are two distinct aspects of survivability that apply at all networking layers.

### IV. INFORMATION ACCESS REQUIREMENT

Does the user have access to the information or service required to complete the task in the presence of failure or attack? For eg. it is possible to replicate service or information and provide them locally when the network gets partitioned. However End-to-end communication should not be mandated in these cases.

### V. END-TO-END COMMUNICATION

On the other hand there are interactive application , inter-personal communication such as voice calls, or dynamically generated information such as current sensor data, which require true end–to–end connectivity. Do existing sessions survive? Can the user create new session to reach the intended communication end- point even in the presence of failures and attacks? This requires that the communication end–point themselves survive and that the communication end–points themselves survive and that the adversary must not be able to permanently partition the network. Furthermore, the adversary must not be able to permanently disable access to required services such as authentication, naming, resource discovery, or routing.

### VI. MOBILE NETWORK SURVIVABILITY

Existing work on survivability in the context of cellular telephone networks concentrates primarily on infrastructure survivability (for e.g. see the outage index metrics (80) and does not consider adversarial attacks. However, they offer some insight on quantifying survivability and the role of network management tools.

Networks are vulnerable during upgrades, especially those involving software. Furthermore, rapid evolution leads to learning – cure problems as well as – over – concentration leads or service into single points of failure. This problem is exacerbated by deficits in network management tools to operate and maintain increasingly complex system. Architectural improvement applicable to mobile include the use of redundant networks
.

### VII. BASE STATION

In more environment, a cell that is geographical region unit is covered by the radio frequency of a base station. Each call is controlled by a BS which has a fixed connection to a BSC (or RNC). In mobile network infrastructure element such as base station controller (BSC), wired links and mobile switch centre (MSC) are employed to provide and maintain essential service. Hence the operation interruption of a network component affects overall or partial network services.

A radiation antenna is classified as omni directional and directional with an ommni directional antenna, a single frequency spreads out in all directions of 360 coverage. A cell is directional antenna with each different set channel.

### VIII. SYSTEM STATE OF BASE STATION

The BS system, including antenna parts cannot provide partial or whole service function for coverage cell when single or more fatal failures occur in the BS system. In this paper, we consider that system failures are caused by key distribution method. For example, by interrupt sequence mishandling, overall system operation falls into failure state because of unanticipated handled interruption to a component of the system.

### IX. PERSONAL AREA NETWORK

A mobile network can have a hierarchical structure e.g. mobile network within another mobile network. This situation is referred to as nested mobile network. A personal area network (PAN) may travel a vehicle, which also contains a mobile network of larger on scale, fig 1 illustrate a simple larger scale. MR-1, MR-2 are attached to their own home links. A wireless personal area network moves as a single unit with one or more mobile routers that connect it to global internet.

## X. MULTI OBJECTIVE OPTIMIZATION (MOO)

An unaccompanied multi objective optimization problem is an example of route optimization for mobile network. Because mobile moves as a single unit with one or more mobile routers that connect it to the global internet[9,11] We defined this problem as
"Minimize" $z = f(x)$
Where $f(x) = (f1(x), f2(x)\ldots fn(x))$
Subject to $x \in x$

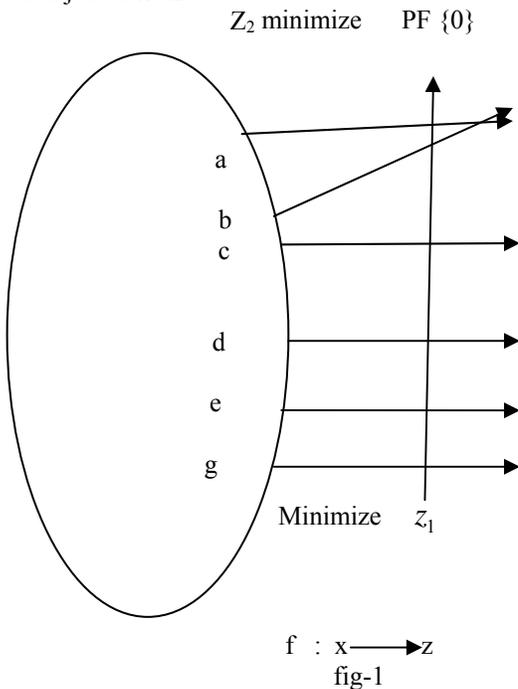

f : x ⟶ z
fig-1

Fig-1 an example of multi objective optimization problem with mobile search space (MSS) x, as vector fitness function f that maps solution in x to objective vector made up of two component (mobile routers) 'costs' $z_1$ and the $z_2$ minimized.
Here if we define
a=mobile router -1
b=mobile router-2
.
.
.
And
1=access router -1
2=access router-2
.
.
.
.

This fig is also defined a routing inefficiency for the traffic management and designed an imported rout optimization schemes for traffic management of mobile networks. Thes concept of traffic management for the network mobility was introduce the signaling over heads of a number of hosts moving as a group as MRs.

## XI. MEMETIC ALGORITHM APPROACH

The impressive record of Memetic Algorithms producing high quality solution in combinalarial optimization and in real – world application (e.g. see page 220[5]) is sometimes cited as a testament of their inherent effectiveness of robustness as black box search. However, since the advent of the no free lunch theorems (6,7,8) we know that MA;S like any other search algorithm, are only reality good to the extent to which they can be "aligned" to the specific features of a route optimization problems in mobile networks. None the less, MAs, like therefore bears evolutionary algorithms (EAs), do have unassailable advantage over other more traditional search techniques: that is their flexibility. This flexibility has important advantage, as has to solve mobile route optimization problems: one is to choose some traditional techniques. And them simplify or otherwise other the problems.

As in any other optimization scenario as route optimization problems, we should know that the out set what is a desirable outcomes; the Memetic Algorithm frame work proposed above required. The operators and procedures be selected based on their current success abs. When a mobile network moves from one place to another it change its point attachment to the internet, which also makes changes to its reach ability and to the internet topology.[9,10,11]

## XII. PERFORMANCE MEASURES IN MAS FOR MOO

If one is developing or using an algorithm for optimization it almost goes without saying that there should be some way to measures its performance. In MOO the situation is the same regarding the time aspect of performance assessment but the quality aspect is clearly more difficult. the extensive array of existing met heuristic, issues and methods reviewed in the section above gives a richer basis from which to design new MAs than do the existing MAs for MOO themselves. In a typical cellular network, the area of coverage is after geographically divided into hexagonal cells. The call is the basic unit of a cellular system.

In recent years, Muscat and Krasnogor have provided a guiding manifesto for putting the "Memetic" back in Memetic algorithm (9,10) advocating.

*Candidate MA framework for MOO*

- MN: = initialize(MN)
- A: = Nondom (MN)

*Algorithm Candidate MA framework for MOO*

- MN: = Initialize (MN)
- MN: = Nondom (MN)
- **while** stop_criterion not satisfied **do**
- **while** stagnation _criterion not satisfied **do**
- SAMN:=SelectFrom(PUA,sel_sc      hed(succ(SEL)))
- SAMN:=Vary(SAMN,var_sched (succ(VAR)))
- SAMN":=LocalSearch(SAMN",I s_sched(succ(LS)))
- MN:=Replace(PUC",rep_sched(s ucc(RED)))
- A:=Reduce(Nandom(AUSAMN),red_sched (succ(RED)))
- **end while**
- MN:=RandomlImmigrants(P,imm_sched(succ(IMM)))
- **return** (A)

Here we represent an Algorithm, we put forward a simple framework that could serve as a guide for making a more Memetic MA for MOO. In line1, MN (Mobile Networks) of solution is initialized. As usual this procedure may be simply random or it may employ some heurists (s).   Line 2 sets the archive A to the no dominated solution from MN. Thereafter, the main loop of the MA begins line 4 sets up an inner loop in which a stagnation criterion is checked. This should be based on some memeplex which monitors progress in diversity, proximity, and /or some other criteria. Line5-9 gives a very high level description of the update of the MN and archive. Five different 'schedulers' are employed, basically corresponding to mating selection, reproduction, lifetime learning, survival selection and update of the archive, respectively. Each scheduler chooses from a memeplex of operators, based on estimates of the current success of those operators. E.g. in line 5, Select from is the operation of mating selection, the domain of which is the union of the MN and archive, and co-domain is a Small Area Mobile Networks (SAMN), the selection is controlled by the scheduler, sel_sched, which use a success measures to choose one operators for the set SEL, of currently available operators for selection. Notice that MN and A are potentially of variable size, in this scheme. In line 11, the MN is updated using some immigration policy to rerelease it from stagnation, the archives of no dominated solution are returned in line 13.

The frame work proposed is rather broad and actually instantiating it requires us to consider how we should resolve many choices, including those considered in the following sections, at the very least?

CONCLUSION

In this paper, we have proposed a scheme for mobile service use of BS system and memetic alogithm. The survavility of Route optimization scheme in nested mobile network modifying the process of Memetic Algorithm . And hence the basic support protocol for survivability of Route optimization scheme for mobile network   needs to be extended with an appropriative route optimization scheme. we proposed scheme can achieve the mobile route optimization environment,  it may get a survivability scheme.

ACKNOWLEDGMENT

The author would like to thank Dr. H.N. Dutta Director, REMTech for his moral support in carrying out the work

AUTHORS PROFILE

Authors Profile .. K K Gautam is the Dean in the Roorkee Engineering & Management Technology Institute, Shamli-247 774, India.

Anurag Rai is the head of the Information Technology Department in College of Engineering Roorkee, Roorkee- 247 667, India.